%Paper: hep-th/9404167
%From: TZANI@vana.physto.se
%Date: Wed, 27 Apr 1994 17:42:45 +0200

\input phyzzx
%\PHYSREV
%\normalspace
%
\Pubnum={USITP-94-06\cr}
\date={April 1994}
\titlepage
\title{The stochastic quantization method for systems with dissipation}
\bigskip
\author{Rodanthy Tzani\footnote\dag
{tzani@vana.physto.se; address after September 1994: Imperial College,
Physics Dept., Prince Consort Road, London SW72BZ}}
\address{Stockholm University, Department of Physics,\break
Box 6730, S-11385 Stockholm, SWEDEN}
\bigskip
\abstract{
The stochastic quantization of dissipative systems is discussed. It is
shown that in order to stochastically quantize a system with dissipation,
one has to restrict the Fourier transform of the space-time variable to
the positive half domain in the complex plane. This breaks the time-reversal
invariance, which manifests in the formulation through the resulting
noninvariant
forms for the propagators. The relation of the stochastic approach
with the Caldeira and Leggett path-integral method is also analyzed.}

\endpage

\def\AP{{\it Ann. Phys.\ }}
\def\PTP{{\it Prog. Theor. Phys.\ }}
\def\NP{{\it Nucl. Phys.\ }}
\def\PL{{\it Phys. Lett.\ }}
\def\PR{{\it Phys. Rev. \ }}
\def\PRD{{\it Phys. Rev. D\ }}
\def\PRA{{\it Phys. Rev. A\ }}
\def\PRL{{\it Phys. Rev. Lett.\ }}

\def\JCP{{\it J. Chem. Phys.\ }}

\def\ZP{{\it Z. Phys.\ }}
\REF\bw{V.~Bargmann and E.P.~Wigner, {\it Proc. Nat. Acad. Sci. U.S.A.}
{\bf 34} (1948) 211.}
\REF\jn{V.P.~Nair and R.~Jackiw, \PRD {\bf 43} (1991) 1993.}
\REF\sh{L.P.S. Singh and C.R. Hagen, \PRD {\bf9} (1974) 898;
C. Fronsdal, \PRD {\bf18} (1978) 3624; J.~Fang and C.~Fronsdal, \PRD
{\bf 18} (1978) 3630; E.S.~Fradkin and M.A. ~Vasiliev, \NP {\bf291}
(1987) 141.}
\REF\dr{W.~Dittrich and M.~Reuter, {\it in ``Classical and Quantum
Dynamics: from Classical Path to Path-Integral}'', Berlin, Germany:
Springer (1992) 341 p.}
\REF\caletal{C.G.~Callan and L.~Thorlacius, \NP {\bf B329} (1990) 117;
C.G.~Callan, A.G.~Felce and D.E.~Freed, \NP {\bf 392} (1993) 551.}
\REF\kos{M.D.~Kostin, \JCP {\bf 57} (1972) 3589.}
\REF\dek{H.~Dekker, \ZP {\bf B21} (1975) 295; \PRA {\bf 16} (1977) 2116.}
\REF\nak{S.~Nakamiya, \PTP {\bf 20} (1958) 948;
I.R.~Senitzky, \PR {\bf 150} (1960) 670; R.~Zwanzig, \JCP {\bf 33}
(1960) 1338; H.~Mori, \PTP {\bf 33} (1965) 423.}
\REF\caleg{A.O.~Caldeira and A.J.~Legget, {\it Physica} {\bf A121}
(1983) 587; \PRL {\bf 46} (1981) 211; \AP {\bf 149} (1983) 374.}
\REF\pol{A. P. ~Polychronakos and R. ~Tzani, \PL {\bf B302} (1993) 255.}
\REF\calan{C. G.~Callan, Jr. and D.~Freed, \NP {\bf 374} (1992) 543.}
\REF\ya{K.~Yasue, \AP {\bf 114} (1978) 479.}
\REF\nel{E.~Nelson, \PR {\bf 150} (1966) 1079.}
\REF\paris{G.~Parisi and Y.S.~Wu, {\it Sci. Sin.} {\bf 24} (1981) 483.}
\REF\com{J.R.~Klauder, {\it Act. Phys. Austr. Suppl.} {\bf XXV} (1983)
251; H.~Gausterer and J.R. ~Klauder, {\it Phys. Rev.} {\bf 33} (1986) 251;
J.~Ambjorn, M.~Flensburg, C. ~Peterson, \NP {\bf 275} (1986) 375;
J.~Ambjorn and S.K.~Yang \NP {\bf 275} (1986) 18; J.~Ambjorn, {\it
In Tbilisi 1986, Proceedings, Quarks '86} 303-315;
A.P.~Polychronakos and R. ~Tzani, \PL {\bf 259B} (1991) 291.}
\REF\na{R. ~Kubo, {\it J. Phys. Soc. Japan} {\bf 12} (1957) 570;
R.~Jackiw and V.P. ~Nair, \PRD {\bf48} (1993) 4991.}
\REF\kel{L.V. Keldysh, {\it Sov. Phys.-JETP} {\bf 20} (1964) 1018;
E.M.~Lifshitz and L.P.~Pitaevskii, {\it Physical Kinetics}
(Pergamon, Oxford,1981).}
\REF\nao{N.~Nakazawa and E. ~Sakane, {\it preprint} NBI-HE-94-19}

%Definitions
\def\half{{1\over2}}
\def\a{\alpha}

\def\g{\gamma}

\def\f{\phi}

\def\t{\tau}
\def\e{\eta}

\def\w{\omega}

The quantization of a classical system may, in general, be hindered
by several obstacles, such as nonrenormalizability, topological
obstructions, anomalies etc. There exists, moreover, a whole class of systems
for which there is {\it no} available method of quantization, namely
non-hamiltonian systems. The quantization of systems with no lagrangian or
hamiltonian formulation, apart from the well-known examples of systems
with dissipation, includes several interesting physical problems, such as
the
Bargmann-Wigner higher spin field equations [\bw] and field theories of
fundamental anyons [\jn]. For the Bargmann-Wigner system of equations
there is no action which directly reproduces this set of equations,
although the quantization for higher spin systems has been done using
more complicated and indirect ways [\sh]. The case of anyonic systems is
known to be unsolved.
The most common example of a non-hamiltonian system is, nevertheless,
the case of a particle on the line with friction. Its equation of motion is
$$
m \ddot x + \g \dot x + {dV \over dx} = 0
\eqn\fric$$
where overdot denotes time derivative and $V(x)$ is the potential
of the particle.

If we assume an action of the form
$$
S= \int f (x, \dot x, t ) \, dt
\eqn\ac$$
for the above equation, the variation of the action gives
$$
{ \delta S \over \delta x } = \int dt \,( {\partial f \over \partial x}
- {\partial \over \partial t} {\partial f \over \partial \dot x } -
{\partial \over \partial x } { \partial f \over \partial \dot x } \dot x
- { \partial ^ 2 f \over \partial \dot x ^ 2 } \ddot x )
\eqn\da$$
Assuming, now, that the second derivative term of the equation
is multiplied by a constant, that is, $ {\partial ^ 2 f \over \partial
\dot x ^ 2 } = m $, the function $f$ can be written as $f= f _ 1 (x, t) +
f _ 2 (x,t) \dot x  + {1\over 2} m \dot x ^ 2 $. Putting this form of
$f$ into \da\ we obtain
$$
{\delta S \over \delta x } = \int ( {\partial f _ 1 \over \partial x }
- {\partial f _ 2 \over \partial t} - m \ddot x ) \, dt
$$
which never contains first derivative terms. Therefore, there is no
action from the variation of which the friction term can be derived
and therefore no canonical structure
and no quantum mechanics for the system \fric.
(Notice that in the search for an action above we do not allow for an equation
of motion which is multiplied by a time-dependent factor.
Such time-dependent factors of the form $ e ^ {f(t)} $ have been
considered [\dr] in the past, but they lead to a time-dependent hamiltonian
and therefore not desirable physics.)

The model \fric\ (and its higher dimensional
versions), apart from its interest as a phenomenological description of
dissipation due to interactions, also has recently found applications in
string theory [\caletal].

There are mainly two approaches that have been used in the literature in
the study of this problem. The first consists of considering the quantization
of these systems as a fundamental physical question and try to find a
direct quantization procedure which classically reduces to the known
dynamics [\kos,\dek].
The second consists in coupling the system with a many degrees of freedom
reservoir. The system plus the reservoir, now, can be described by a
hamiltonian and the dissipation is due to the interaction of the system
with the reservoir. In this case the problem is reduced to finding a
way to ``integrate out'' the extra degrees of freedom of the reservoir
in order to finally obtain a quantum mechanical description of the original
system [\nak--\caleg]. The first approach is more fundamental and mathematical
while the second is more physical.

Using the second approach in the study of tunneling in dissipative
systems, Caldeira and Leggett [\caleg]
have derived an effective action for
\fric\ in the euclidean time. In their path-integral approach the term
which classically produces the friction force corresponds to a non-local
term in the action. Their effective action is given by
$$
S_{(eff)} (x(t))= \int _0 ^ \t [ {1\over2} m \dot x ^ 2 + V(x) ] dt
+ { \g \over 4 \pi} \int _ { - \infty} ^ {\infty} \int _ 0 ^ \t
dt d t ^ \prime { [ x(t) - x(t ^ \prime) ]  \over (t -t ^ \prime ) ^2 } ^ 2
\eqn\cl$$
The approach of Caldeira and Leggett has, however, some disadvantages. Coupling
the system to an infinite set of oscillators and then integrating them
out, leads to a state where the fourth moment of the field variable diverges.
This is due to the zero point motion of the set of oscillators, which
perturb the system in a substantial way [\pol]. Moreover, it is not
clear that this approach is equivalent to a fundamental
quantization procedure of dissipative systems. Nevertheless, this action
has been used in the literature in a similar context in order to describe
the quantum effect of friction [\calan].

The first approach was followed by Kostin [\kos] who has proposed a modified
Schr\"odinger equation for the model \fric\ in which the friction
is reproduced through a wavefunction-dependent potential. This approach
exhibits also some drowbacks: it violates the
superposition principle, and has stationary states in which the
energy does not dissipate. (The same equation was rederived by Yasue [\ya]
using Nelson's stochastic quantization scheme [\nel].)
Although we do not expect the quantum mechanics of dissipative systems
to have the same properties as the standard ones, the fact that one obtains
stationary non-dissipative states for these systems is counter-intutive.
Due to these rather unphysical properties Kostin's equation has not been
used extensively in the literature.

This last approach of a modified Schr\"odinger equation has been extended
further by Polychronakos and the author in a previous paper [\pol] by
allowing a wavefunction-dependent first-derivative term. This study
resulted in
a parametric non-linear Schr\"odinger equation for the particle with
friction which reproduces Kostin's equation for a specific value of the
parameter. This family of equations avoids some of the problems that
Kostin's equation exhibited, such as the stationary states.

In the present paper, taking the point of view of fundamentally quantizing
the friction system, we study the stochastic quantization approach of
equation \fric. Stochastic quantization [\paris] has the advantage that
it relies only on the classical equation of motion.
It essentially consists of reproducing the
weighting factor $ \exp (-S)$ of the euclidean path-integral as a limiting
probability density of a particular stochastic process. The stochastic
process which reproduces the path-integral results for bosonic
fields is given by the following Langevin equation
$$
{\partial \f \over \partial \t} = - { \delta S \over \delta \f } + \e ,
\eqn\le$$
where ${\delta S \over \delta \f} $ is the classical equation of motion,
$\t$ is the stochastic time and
$\e$ is a white noise over spacetime and $\t$. In the stochastic limit
$\t \rightarrow \infty$ the fields attain their equilibrium distribution
with probability density equal to $ \exp (-S) $.

Since the stochastic quantization method is based on the equation
of motion, one would naively expect to overcome the problem of hamiltonian
formulation of these systems and be able to quantize them
directly using stochastic quantization method.
In this work, we presend some peculiarities related with the stochastic
quantization approach of dissipative systems.

We consider the simplest example of a particle moving on the line with
friction. Its equation of motion is given by \fric\
$$
m \ddot x + 2 \g \dot x + k x= 0 ,
\eqn\2$$
where we have multiplied the friction term by 2 for later convenience and have
assumed the harmonic oscillator potential $V(x) = {1 \over 2} k x^2$.
The Langevin corresponding to equation \2\ is given by the
following euclidean time equation
$$
{\partial x (t , \t) \over \partial \t}  = m \ddot x (t, \t)  -2i\g \dot x
(t , \t) - k x (t, \t) +  \e (t, \t)
\eqn\l$$
where
$$
< \e (t ,\t )> _ \e \, =\, 0
\,\,\,\,\,{\rm and} \,\,\,\,\,
< \e (t , \t ) \e ( t ^\prime , \t ^ \prime ) > _ \e \,=\,
2 \delta (t - t ^ \prime ) \delta ( \t - \t ^ \prime ) \, .
$$

The first thing to notice about this Langevin is that it is complex.
Since the friction term is first order in time derivative, after the
Wick rotation to euclidean time this term is imaginary in the expression
of $\delta S \over \delta x$.
Therefore, one faces the problem of stochastically quantizing complex
actions [\com] which has not been solved in its generality. In this
particular case, however, since the equation of motion is linear in
the $x$ variable, the problem can be solved. The solution
of the Langevin equation is easily obtained in the Fourier space. The Fourier
transform of \l\ is
$$
{\partial \tilde x (\w ,\t) \over \partial \t } = (-m \w ^2 + 2 \g \w -k )
\tilde x  (\w, \t) + \tilde \e (\w ,\t)
\eqn\fl$$
where $\tilde x(\w ,\t) $ and $\tilde \e (\w ,\t) $ are the Fourier transforms
of $x(t, \t) $ and $ \e (t, \t)$ correspondingly.
(Notice that for overcritical damping, that is $\g > \g _ c = \sqrt {km}$,
\fl\ will develop instabilities for values of $\w$ such that $
{\g \over m } - \sqrt { {\g ^ 2 \over m ^ 2 } - {k \over m} } < \w <
{\g \over m } + \sqrt { {\g ^ 2 \over m ^ 2 } - {k \over m} }$.)

Now, in the Fourier space
the Langevin becomes real and it naively seems that one avoids the complex
action problem by going to Fourier space. This apparent resolution is
illusionary, however, and is due to the Fourier transform of first order
derivatives ($\tilde \e (\w, \t)$ is complex). As we will see in what follows,
the problem
related with the first order in time derivative term manifests itself in the
impossibility to obtain $\g$-dependent results in this approach.
The solution of the last equation is given by
$$
\tilde x (\w ,\t) = \int _ 0 ^ {\t} d \t ^ \prime \, \tilde \e (\w ,\t
^ \prime) \, e ^ {(k -2\g \w + m  \w ^ 2 ) ( \t ^ \prime - \t)}
\eqn\sl$$
where now the $ \e $-average in the Fourier space is
$$
< \tilde \e ( \w ,\t ) \tilde \e ( \w _ 1 , \t _ 1) > _ \e = 2 \delta ( \w +
\w _ 1 ) \delta (\t - \t _1 ) \,.
\eqn\ev$$

The interesting quantities to be computed are correlation functions of the
variable $ x(t)$.
In the Fourier space they are
$$
< \tilde x (\w  , \t )  \tilde x (\w _ 1 , \t _ 1 ) >_ \e
$$
$$
= \, < \int _ 0 ^ \t d \t ^ \prime  \int _ 0 ^ {\t _ 1} d \t ^ {\prime \prime}
\tilde \e (\w , \t ^ \prime) e ^ {(k -2 \g \w + m \w ^ 2) (\t ^ \prime -
\t )} \tilde \e ( \w _ 1 , \t ^ {\prime \prime} ) e ^ { (k -2 \g \w _ 1
+ m \w _ 1 ^ 2) ( \t ^ {\prime \prime } - \t _ 1 ) } > _ \e
$$
$$
= \,2 \delta ( \w + \w _ 1 ) { 1 \over { (k -2 \g \w + m \w ^ 2 )\, +\,
 ( k -2 \g \w _ 1  + m \w _ 1 ^ 2 ) } }
\eqn\cw$$
where we have used the relation \ev\ and have taken the limit $\t= \t _1
\rightarrow \infty $ in order to obtain the last result.

Next, we Fourier transform the last two-point function back to the original
space.
Because of the form of the white noise average in the Fourier
space, namely, given as a delta function of the sum of the $\w$'s in
\ev, all linear terms in $\w$ cancel after the integration over $\w$.
We obtain
$$
< x (t)  x(t ^\prime ) > \,=\, {1 \over 2 \pi} \int d \w { 1 \over
m \w ^ 2 + k } e ^ { i \w (t - t ^ \prime) }
\eqn\co$$
This integral is calculated by contour integration.
The poles ($ \w = \pm i \sqrt {k \over m} $) lie on the imaginary axis and
for $ t > t^\prime$ we close the contour in the upper half complex plane
while for $t <t ^ \prime $ we close the contour in the lower half plane.
The result obtained is
$$
<x (t) x(t ^ \prime ) > \,=\,{1 \over 2 \sqrt {km}} [ \theta (t - t ^ \prime)
e ^ {- \sqrt {{k \over m }}
(t - t ^ \prime) } + \theta (t ^ \prime -t ) e ^ { \sqrt {{k \over m}}
(t - t ^\prime) } ]
\eqn\ro$$
which gives
$$
<x ^ 2(t) > \,=\,{1 \over  \sqrt {km} }
\eqn\cf$$
for the equal time correlation function of the variable $ x(t) $.
These results
coincide with the results one obtains from the quantization of this
system for the $ \g = 0 $ case.
Therefore, the naive application of stochastic quantization for this
system gives results
for the correlation functions that do not depend on the friction term.
(It is a straightforward calculation to show that none of the higher order
correlation functions depend on the friction parameter.)

The meaning of the last result is the following:
adding a friction term in the equation of motion is equivalent to adding
a ``curl'' term in the Fokker-Planck hamiltonian. Indeed, the Fokker-Planck
equation for the system is
$$
\int dt \, \left[{ \delta \over \delta x (t) } \left( {\delta \over \delta
x(t) } + {\delta S _ 0
\over \delta x(t) } + F \right) \right] P \,=\, - { \partial P \over
\partial \t}
\eqn\fp$$
where ${ \delta S _ 0 \over \delta x(t)} $ is the part of the equation of
motion which is derived from an action and $F$ is the friction force, given
by $ - 2i \g \dot x (t) $ in our case.
Because of the form of the Fokker-Planck equation we can identify the
probability current as $ J _ x = - ( {\delta \over \delta x(t) }  +
{\delta S_ 0 \over \delta x (t) } + F ) P $ and $P$ is the probability
density, which satisfy the probability conservation equation
$$
\int dt { \delta J _ x \over \delta x (t) } + {\partial P \over \partial
\t } = 0 \,.
\eqn\pc$$
Then one can show that the part of the current $J_f$
which comes from the friction force is divergenceless.
Indeed, the Fokker-Planck equation in the equilibrium limit is satisfied
by $ P = e ^ {- S _ 0} $, and
$$\int dt {\delta J_f \over \delta x (t)} =
\int dt {\delta \over \delta x (t)} (F e^{-S _0 })= 0 \, .
\eqn\Jf$$
(The last relation is easily shown in the $\w$ plane and
it is based on the fact that the range of integration in the $\w$ variable
is from ${-\infty}$ to $\infty$ while the integrand is an odd function
of $\w$.) That is, $ {\vec \nabla } \cdot {\vec J _ f } = 0$.
%(This, then, means that the current $ \vec J _ f $ can be
%expressed as
%the ``curl'' of some vector $A$; namely, $ {\vec J _f }= {\vec \del } x
%{\vec A} $.)
Therefore, the friction force produces currents that do not change the
probability density, which corresponds to the fact that they do not alter
the physical results.
%Notice that in the case of systems with an action,
%the probability current at $\tau=\infty$ vanishes and therefore the ground
%state is static, while in this case, where there is a nonvanishing current,
%it is merely stationary.

One, however, can approach the problem in the following way:
let's view $x(t)$ as a $0+1$-dimensional field theory. Then, for the case
$\g =0 $, $x(t)$ is real and it describes only particles.
(Particles here correspond to energy quanta.) In this
description, in the Feynman picture, positive frequency $\w>0$ which
propagates forward in time
and negative frequency $\w<0$ which propagates backward in time corresponds
to the same particle state. Since
friction breaks the time-reversal invariance, this is not a good description
for the $\g \neq 0 $ case. There exists, however, an equivalent description
for particles in the frictionless case. This can be obtained
by considering a complex field but keeping only ${\it positive}$
frequencies and ${\it forward}$ propagation in time.
This is, then, a more appropriate way to incorporate
friction. In order to achieve it, we make the assumption that the
field $x(t)$ in our Langevin equation is complex with only positive
frequencies, that is, an analytic function in $t$ in the upper half
complex plane.

Given a real function $ x(t)$ we can always
construct a function analytic in the upper half plane
(in general complex), with the use of the Hilbert transform of the function,
as follows
$$
x _ \a (t) =  x (t) + { i \over \pi }P  \int d t ^ \prime
{ x (t ^ \prime) \over t - t ^ \prime }
 \equiv  x(t) +  i y (t)
\eqn\an$$
where by $P $ in front of the integral we mean principal value.
With this $x _ \a(t)$, $ \tilde x _ \a (\w) = 0 $ for any $\w <0$.
We assume that the Langevin equation \l\ is satisfied by the function
$x _ \alpha (t, \t) $. Then, in order for the Langevin to be consistent,
the white noise function should obey the same analyticity property.
That is
$$
\e _ \a (t) = \e (t) + {i \over  \pi } P \int d t ^ \prime
{ \e ( t ^ \prime ) \over t - t ^ \prime  } \equiv \e (t) + i \e _ i (t)
\eqn\el$$

In order to calculate correlation functions in this framework we must
remember that our Langevin describes only particles and not antiparticles.
Then, for $t >t ^ \prime$
the appropriate correlation function to be computed is
$$
< x _ \a (t ) x ^ *  _ \a (t ^ \prime) >
\eqn\pc$$
This gives the propagator for the particles moving forward in time from
$ t ^ \prime$ to $t$.
(The propagator for the antiparticles is given by the complex conjugate
of this quantity.)

Again it is convenient
to go to the Fourier space. The Langevin is
given by \fl\ where $\tilde x (\w, \t)$ and $\tilde \e (\w, \t)$ have been
replaced by $\tilde x _ \a (\w, \t) $ and $ \tilde \e _ \a (\w, \t) $
correspondingly. Since the white noise is an analytic function in the
$t$-plane it
satisfies the condition $ \tilde \e _ \a (\w, \t) =0 $ for any $\w <0 $ in
the $\w$-plane. Then the correlation functions for $ \tilde \e _ \a $
obtained from \ev\ are
$$
\eqalign {
< \tilde \e _ \a (\w , \t) \tilde \e _ \a ( \w ^ \prime , \t ^ \prime ) >
\,&=\, 0 \cr
< \tilde \e _ \a (\w ,\t) \tilde \e ^ * _ \a (\w ^ \prime , \t ^ \prime ) >
&= 8 \, \delta ( \w - \w ^ \prime ) \delta ( \t -\t ^ \prime) \cr }
\eqn\dw$$
for both $ \w$ and $ \w ^ \prime$ greater than zero.

Following the same procedure as before and
using the last relations for the $\e$-average we obtain the following
expression for the two-point function of the analytic $ x _ \a (t)$
$$
<x _ \a(t) x ^ * _ \a (t ^ \prime) > =
{4 \over 2 \pi} \int _ 0 ^ {\infty}
d \w { 1 \over m \w ^ 2 - 2 \g \w + k } e ^ { i \w (t - t ^ \prime) }
\eqn\ie$$
which depends explicitly on the friction parameter.

Next, we compute the last integral. We are interested only in the case
that $t>t ^ \prime$.
For $ t > t ^ \prime$ we can evaluate the integral by deforming the contour
into the upper half $\w$ plane.
The integrand has poles at
$\w = {\g \over m } \pm i \alpha $
where $ \alpha \equiv \sqrt { {k \over m } - {\g ^2 \over m ^2 }}$
and we have assumed that $ \g ^ 2 < km $ .
If we further make the assumption that we are close to critical damping,
that is, $\g^2 -km $ is a very small quantity, the poles will lie
close to the real axis.
We deform the contour as shown in fig. 1. Application of  Cauchy's theorem
around the contour C gives the following expression for
the above integral
$$
 { 4 \over 2 \pi } \int _ \epsilon ^ {\epsilon +i \infty }
d \w {1 \over m \w ^ 2 - 2 \g \w + k } e ^ {i \w (t - t ^ \prime) }
+{ 4\over 2m \a } e ^ {({ i\g \over m } - \a ) (t - t ^ \prime ) }
\eqn\in$$

Then if we assume that $|t| {\g \over m } >> 1 $ and  $ |t| \a \leq 1 $ we
can show that the integral in the above
expression is negligible and we obtain the following result
$$
<x _ \a(t) x ^ * _ \a (t ^ \prime) > =
{2 \over \sqrt {km - \g ^ 2} } e ^ {(i {\g \over m}
 - \sqrt {{k \over m} - {\g^2 \over m^2}}) (t - t ^ \prime) }
\eqn\ac$$

The last expression gives the propagator for the problem of a particle
with friction. In Minkowski space this propagator is not unitary due
to the reality of the term proportional to $ {\g \over m}$ in the exponent.
This means that there is particle loss in this picture,
which corresponds to the decay of energy due to dissipation, since particles
correspond to energy quanta.
Indeed, from the classical equation of motion, the change in the energy
($E$) is given by
$$
\dot E = -2 \g \dot x ^ 2
\eqn\de$$
Choosing, now, our state to be an almost ``energy eigenstate'' (which can
be the case for $\g<<1 $) we obtain
$$
< \dot E > = -2 \g < \dot x ^ 2> = -{2 \g \over m } <E>
\eqn\ox$$
for the decay of the energy due to the friction. This is the same as
the decay rate we obtain for the energy from the result \ac.

Finally, one could solve the
Langevin equation \l\ by separating its real and imaginary parts.
We obtain the following coupled equations for $x(t)$ and $ y(t)$.
$$
{ \partial  x (t , \t) \over \partial \t} = m \ddot x (t, \t) + 2 \g \dot y
(t, \t ) - k x(t, \t) +  \e (t, \t)
$$
and
$$
{ \partial y (t , \t) \over \partial \t } = m \ddot y (t, \t) - 2 \g \dot x
(t , \t) - k y (t, \t) + \e _ i (t , \t)
\eqn\ce$$
where $\dot y (t) = - {1\over  \pi} P
\int {x(t ^ \prime ) \over (t - t ^ \prime ) ^ 2 } dt ^ \prime $ and
$ \ddot y (t) = {2 \over \pi} P \int {x (t ^ \prime ) \over (t - t ^\prime
) ^ 3}  d t ^ \prime $ and by principal value, here, we mean
$$
P \int { x (t ^
\prime)  \over (t - t^ \prime) ^ n } d t ^ \prime = {1 \over 2} \left[ \int
d t ^ \prime { x(t ^ \prime) \over (t -t ^ \prime + i \epsilon ) ^ n}
+ \int d t ^ \prime { x (t ^ \prime ) \over (t - t ^ \prime - i \epsilon )
^ n }  \right] .
\eqn\pv$$
The stochastic process is, then, defined by the equations \ce.

The corresponding equation for $\tilde x (\w, \t) $ is
$$
{\partial \tilde x ( \w, \t) \over \partial \t } = ( -m \w ^ 2 + 2
 \g |\w| - k)
 \tilde x (\w, \t) +  \tilde \e ( \w , \t)
\eqn\la$$
The last equation coincides with the original Langevin \fl\ after the
substitution of $\w$ by $ |\w|$ and
gives results for the correlation functions of $ x(t)$ which depend on the
friction parameter.

Indeed, with this Langevin the two-point function
is expressed as an integral over $\w $ in the following way
$$
< x(t)   x (t ^ \prime ) > = {1 \over 2 \pi} \int _ {-\infty} ^ {\infty}
 d \w { 1 \over
 m \w ^ 2 - 2 \g |\w| + k }  e ^ { i \w (t -t ^ \prime) }
\eqn\sc$$

The calculation of this integral follows similar lines as the previous one
and the result for the correlation function is given by ${\half}$
of the real part of the
expression \ac. (Remember that $x(t)$ here is the real part of $x _ \a (t)$
and that $\tilde y (\w) = -i \, sgn(\w)\, \tilde x (\w)$,
where by $sgn(\w)$ we denote
the sign of $\w$.) This expression for $ \g =0$ reproduces the standard
harmonic oscillator result \co.
The full expression \ac\ can be recovered from \sc\ by using dispersion
relations.

In what follows, we compare the stochastic approach with
the path-integral approach of Caldeira and Leggett. We find that
the stochastic quantization one obtains starting from the
Langevin given by equations \ce\ is equivalent to the
stochastic quantization one would obtain starting from the effective
action of Caldeira and Leggett.

In order to see this, it is convenient to rewrite the non-local term of the
action \cl\ with the use of the identity $ x(t) x (t ^ \prime) \equiv
{1\over 2} [x ^ 2 (t) + x ^2 (t ^ \prime) ] - {1 \over 2} [ x (t) - x
(t ^ \prime) ]^ 2 $.
The effective action can be expressed as
$$
S _ {(eff)} \,=\, \int _ 0 ^ {\t} [{1\over 2}m \dot x ^ 2 + V (x) ] dt
- { \g \over 2 \pi } P \int _ {- \infty} ^ {\infty} \int _ 0 ^ \t
dt dt^ \prime { x (t) x(t ^ \prime) \over (t-t ^ \prime)  ^ 2 } +
{\g \over 4 \pi } P \int _ {- \infty} ^ {\infty} \int _ 0 ^ {\t }
d t d t ^ \prime { x ^ 2 (t) + x ^ 2 (t ^ \prime)  \over (t - t ^ \prime)
^ 2 }
\eqn\ca$$
where the principal value in the last integrals is necessary in order to
take care of the infinity at $ t =t ^ \prime$.
The last modification one has to do is to extend the limits of integration
in \ca\ from ${ -\infty}$ to $ \infty$ [\caleg] . It can be easily, then, shown
that the last integral vanishes and the variation with
respect to $x(t)$ of the remaining $\g$-dependent term in \ca\ gives
$$
- \, { \g  \over  \pi} P \int d t ^ \prime { x(t ^ \prime) \over
(t - t ^ \prime ) ^ 2 }
\eqn\cag$$
which coincides with the friction term of the first of the equations \ce,
was not for the factor of $2$ we have multiplied our friction term.

Concluding, we like to emphasize that it seems peculiar that the stochastic
quantization of our dissipative model is achieved by restricting the Fourier
transform of the space-time variable to the positive half domain in the
complex plane. The physics, however, dictates this choise as it is argued
in the text. Choosing the solution of the Langevin equation to be
a complex function is natural, since in our original Langevin $x(t)$
was complex. The extra restriction, however, to be analytic in $t$
in the upper
half plane stems from the fact that the Langevin should break the
time-reversal invariance in order to describe friction.
Notice at this point that if one complexifies both $x(t)$ and $\eta (t)$
in the Langevin one obtains doubling in the degrees of freedom of the
problem. This doubling is eliminated after restricting the frequencies in
the positive half domain in the complex plane.

It is interesting, nevertheless, to connect our approach with what is
known about this problem. Due to the time-reversal non-invariance of the
dissipative processes, retarded Green's functions
are more appropriate to use in order to describe them than
the symmetric time-ordered ones [\na]. The reason is that,
since the system looses energy irreversibly, a function with only
forward propagation in time can describe it. The standard stochastic
approach, however, gives the Feynman propagator, which is symmetric
under time reversal. Indeed, our two-point function \ro\ obtained
by straightforward application of stochastic quantization does not
break time-reversal and does not describe dissipation. Formulating, on the
other hand, the problem such that it involves fields which are analytic in
the upper half plane (or equivalently fields that contain only positive
frequencies) we have broken time-reversal in such a way that when time
runs from ${-\infty}$ to ${\infty}$
these fields are appropriate to describe the forward propagation in time.
The complex conjugate of these fields gives the antiparticle propagation,
which can be equivalently obtained by performing a time reversal operation
in the original fields. The exact relation, however, of our approach with
the retarded Green's function approach, or equivalently with the
Keldysh formulation [\na --\kel] for nonequilibrium phenomena in quantum
field theory, is a subject for further investigation.

It is also interesting to notice that the fact that the stochastic
approach with the $\w$ restricted in the positive half plane gives
friction-dependent results is related to the divergence of the
corresponding probability current as explained earlier. Indeed,
in this case, due to
the fact that the range of integration of $\w$ is from $0$ to ${\infty}$,
the calculation of the divergence of the $ J _ f$ current
leads to a nonzero result. Therefore, the friction
currents change the probability density in this case, as they should.

Finally, in general, the stochastic quantization of dissipative systems will
work in similar patterns; that is, one has to complexify each real coordinate
of the system and keep only positive frequencies.

\bigskip
The material of this paper was presented at the ``Lattice Field Theory
Workshop'', Vienna, June 11-12, 1993
and at the ``IV International conference on mathematical physiscs,
string theory and quantum gravity'', Alushta, 13-24 June, 1993.

After the completion of this work, preprint [\nao] appeared where
the prescription $ \w \rightarrow |\w|$ is used in the finite temperature
stochastic quantization of the system.

\medskip
\ack{I would like to thank I. Bengtsson for pointing out to me references [3],
H. Hansson for discussions, V.P. Nair for useful comments and a critical
reading of the manuscript, and A.P. Polychronakos for many enlightening
discussions and for a critical reading of the manuscrtipt.}

\refout
\end